\documentclass[journal=mamobx,manuscript=article]{achemso}
\setkeys{acs}{articletitle = true}

\usepackage{graphicx}% Include figure files
\usepackage{dcolumn}% Align table columns on decimal point
\usepackage{bm}% bold math
\usepackage{hyperref}% add hypertext capabilities
\usepackage[dvipsnames]{xcolor}
\usepackage{amssymb}

\usepackage[version=3]{mhchem} % Formula subscripts using \ce{}

\author{Davide Breoni}
 \affiliation{Department of Physics
, Università di Trento, Via Sommarive 14, I-38123 Trento, Italy }%Lines break automatically or can be forced with \\
 \alsoaffiliation{INFN-TIFPA, Trento Institute for Fundamental Physics and Applications, I-38123 Trento, Italy}
  \email{davide.breoni@unitn.it}
\author{Emanuele Locatelli}%
\affiliation{Department of Physics and Astronomy, University of Padova, Via Marzolo 8, I-35131 Padova, Italy}
\alsoaffiliation{INFN, Sezione di Padova, Via Marzolo 8, I-35131 Padova, Italy}
\author{Luca Tubiana}
 \affiliation{Department of Physics
, Università di Trento, Via Sommarive 14, I-38123 Trento, Italy }%Lines break automatically or can be forced with \\
 \alsoaffiliation{INFN-TIFPA, Trento Institute for Fundamental Physics and Applications, I-38123 Trento, Italy}

\title{Effects of knotting on the collapse of active ring polymers}

\abbreviations{IR,NMR,UV}

\begin{document}

\begin{tocentry}

\includegraphics[width=1.00\textwidth]{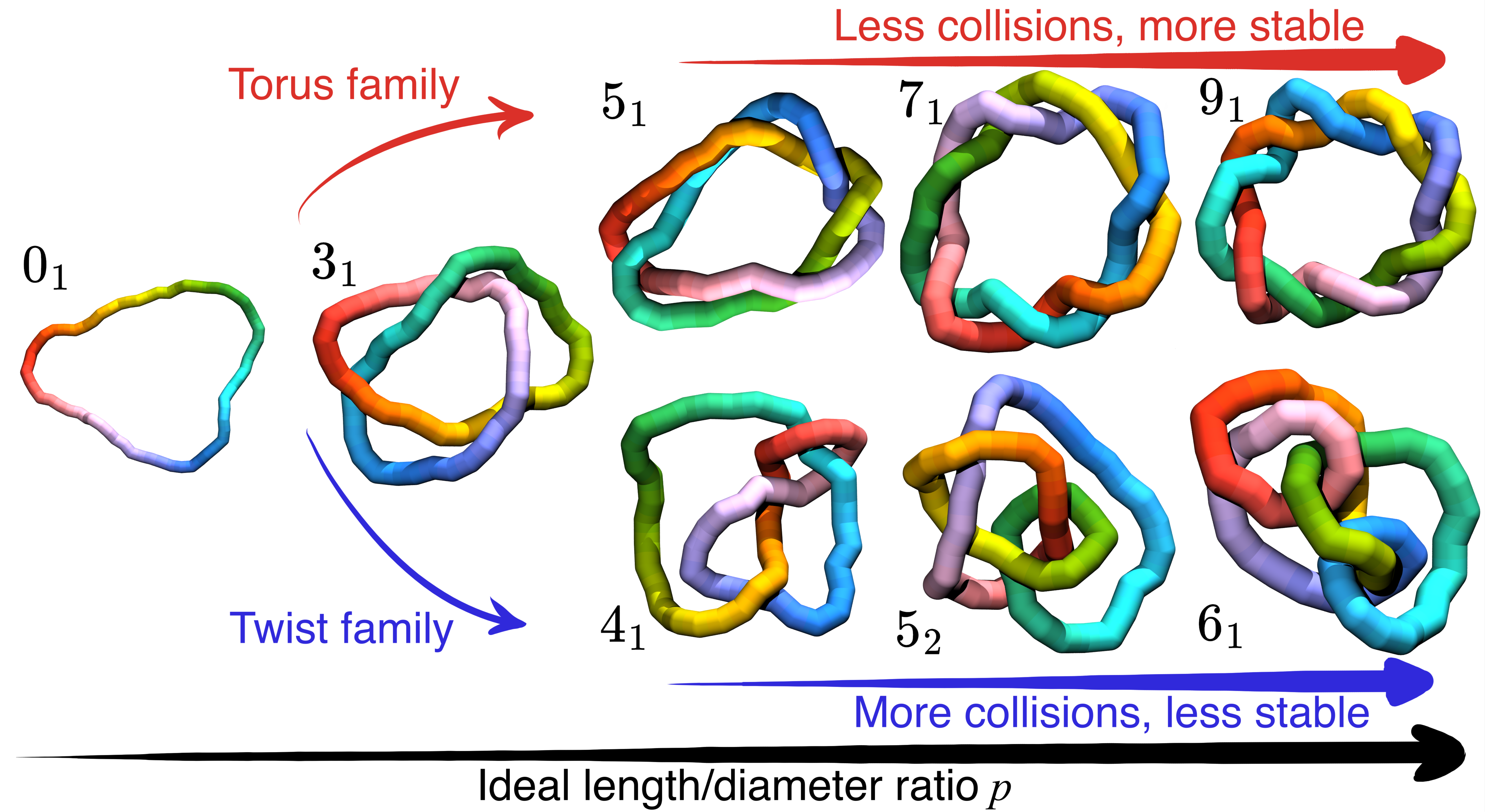}
%\\
%\\
%For Table of Contents Use Only

\end{tocentry}

\begin{abstract}
  We use numerical simulations to study tangentially active flexible ring polymers with different knot topologies. Simple, unknotted active rings display a transition from an extended phase to a collapsed one upon increasing the degree of polymerization. We find that topology has a significant effect on the polymer size at which the collapse takes place, with twist knots collapsing earlier than torus knots. Increasing knot complexity further accentuates this difference, as the collapse point of torus knots grows linearly with the minimum crossing number of the knot while that of twist knots shrinks, eventually canceling the actively stretched regime altogether. This behavior is a consequence of the ordered configuration of torus knots in their stretched active state, featuring an effective alignment for non-neighboring bonds which increases with the minimal crossing number. Twist knots do not feature ordered configurations or bond alignment, increasing the likelihood of collisions, leading to collapse. These results show that topology yields a degree of control on the properties of active ring polymers, and can be used to tune them. At the same time, they suggest that activity might introduce a bias for torus knots, as complex twist knots cannot be formed in extended active polymers.

\end{abstract}

%% QUESTIONS ASKED IN VENICE
%% Q1. Beautiful talk, but you should consider hydrodynamics. --> How do we justify ignoring it? What do we think will happen when it is introduced?
%% R1. Ask Ema. Perhaps we can justify it by saying we use low activities? Do we? If activity was to be included, I would expect the same behaviour for torus knots though, as they will simply rotate. I thus expect the local environment around the beads to remain on average the same. For twist knots I don't know, but I can't imagine the inclusion of hydrodynamics reducing the probability of collisions.

%% Q2. What happens if you consider chirality?
%% R2 Apparently nothing. Add to the SI.

%% Q3. What happens for T3 torus knots?
%% R3. Perhaps we can add something in the SI pointing to the necessity of 
%% a larger study
\section{Introduction}
Polymers can attain a wide range of different topologies\cite{tubiana_topology_2024}, from unknotted rings~\cite{haque_synthesis_2020} to knots and links~\cite{orlandini2021topological}, and from interconnected networks in melts~\cite{michieletto2016tree, bonato2022topological} to mechanically interlocked molecules~\cite{hart_material_2021}. For example,  actin has been tied into knots in vitro and knots have also been found in a small but important fraction of proteins~\cite{arai_tying_1999,jamroz_knotprot_2015,majumder_knots_2021}. DNA can be organized in supramolecular structures that present various degrees of topological complexity, from chromatin loop networks during interphase~\cite{bonato2024topological} to looped bottlebrush-like shapes during duplication~\cite{harju_physical_2025}, and from Olympic networks, typical of the Kinetoplast DNA (kDNA, the mitochondrial DNA of the parasites Trypanosomatids~\cite{ramakrishnan_single-molecule_2024}) to  knots~\cite{arsuaga_dna_2005,valdes_dna_2018}. In fact, knots become extremely likely in long and confined polymers, including genomic DNA~\cite{liu_knotted_1981,tubiana_spontaneous_2013,plesa_direct_2016}, impacting its sequencing and behavior~\cite{suma_pore_2017,klotz_dynamics_2017,ma_diffusion_2021}. More generally, knots have a large impact on polymers' configurations~\cite{rawdon_effect_2008,millett_effect_2009,lang_effect_2012,baiesi_knotted_2014}, and show a complex dynamics~\cite{iwata_topological_1991,orlandini_how_2016,weiss_hydrodynamics_2019,tagliabue_tuning_2022} inclusive of inter-knots interactions~\cite{dai_effects_2016,ruskova_controlling_2025}.

A typical property of biological systems is \textit{activity}, that is the ability of turning external energy into directed motion, a mechanism by which individual agents inject energy into the system, driving it out-of-equilibrium. It is responsible for a plethora of phenomena, such as self-organization\cite{hagan2016, needleman_active_2017}, collective motion and flocking~\cite{vicsek2012collective, cavagna_bird_2014}, spontaneous flow~\cite{marchetti2013hydrodynamics} and clustering in absence of attractive forces (also known as Motility-Induced Phase Separation, or MIPS)~\cite{buttinoni_dynamical_2013,cates_motility-induced_2015}.
Active polymers and filaments~\cite{winkler_physics_2020}, the focus of much theoretical and experimental interest in the past few years, present an even richer phenomenology. Besides the aforementioned collective motion~\cite{sanchez_cilia-like_2011,sumino_large-scale_2012,ozkan2021collective}, self-organization~\cite{miranda2023self,patil2023ultrafast,faluweki2023active} and clustering~\cite{deblais_phase_2020}, active filaments display pattern formation~\cite{schaller2010polar}, enhanced long-time diffusion constant~\cite{bianco_globulelike_2018, tejedor_reptation_2019}, anomalous rheological properties~\cite{deblais_rheology_2020} and giant melt elasticity~\cite{breoni_giant_2025} 
%complex configurations~\cite{bianco_globulelike_2018,ubertini_universal_2024}, and a collapse transition at high degrees of polymerization~\cite{locatelli_activity-induced_2021} 
that put them at the forefront of the research for dynamically tunable nanomaterials~\cite{koenderink_active_2009,bertrand_active_2012,li_macroscopic_2015, burla_mechanical_2019, sciortino2025active} as well as for advanced macroscopic soft robots\cite{deblais2023worm}. These nanomaterials are, for the most part, based on biopolymers such as DNA~\cite{bertrand_active_2012,li_macroscopic_2015}, microtubules~\cite{sciortino2025active} and actin~\cite{koenderink_active_2009}, that are among the most relevant polymers in nature. Activity comes from molecular motors that, for actin and the other cytoskeletal filaments, allow to provide the cell with rigidity and motility, while, for DNA, play crucial roles %DNA actively moves as proteins move along its backbone 
during replication, transcription and protein production~\cite{alberts_essential_2015}. %Active filaments are relevant across scales, with examples ranging from cilia and flagella~\cite{balin_biopolymer_2017,beeby_propulsive_2020}, to cyanobacteria~\cite{faluweki2023active,kurjahn2024collective,rosko2025cellular} and other micro-organisms~\cite{patra2022collective}, or C.elegans~\cite{sugi2019c} and larger worms~\cite{deblais_phase_2020}. In addition, synthetic active polymers have been engineered both at the micro-scale, with active polymers made of colloids or droplets~\cite{yan2016reconfiguring,lowen_active_2018,kumar2024emergent}, artificial cilia and flagella~\cite{van_oosten_printed_2009,vutukuri2017rational} as well as at the macro-scale with, as already mentioned, soft robots~\cite{ozkan2021collective, becker2022active, deblais2023worm}.\\

%In addition to the relevance of activity and topology in biological polymeric systems and their promising applications% in the engineering of novel materials, 
The study of active polymers has recently revealed interesting, unexpected, interplay with topology.
Dense suspensions of active diblock ring polymers, where activity is realized as an additional higher temperature, display glassy behavior\cite{smrek2020active}, originating from a network of entanglement called deadlocks~\cite{micheletti_topology-based_2024}. 
In the case of tangential activity, where the self-propulsion of each monomer follows the local backbone conformation, 
it was shown that active polymers are much more efficient in forming knots than their passive counterpart~\cite{li_activity-driven_2024,vatin_upsurge_2025} and are consequently less prone to adsorption~\cite{shen_knotting_2025}. Even in the case of simple, unknotted ring polymers, simulations showed a swelling-collapse transition at infinite dilution, driven by activity~\cite{locatelli_activity-induced_2021}, while in the semidilute regime swollen rings display clustering features~\cite{miranda2023self}. 

In this work we employ numerical simulations performed over a wide range of polymerization degrees to  %instead on already
systematically characterize how knot topology affects the properties of active polymer rings, and in particular the swelling-collapse transition observed in Ref.~\cite{locatelli_activity-induced_2021}. 
%More specifically we employ numerical simulations to explore their activity-induced collapse~\cite{locatelli_activity-induced_2021} as a function of their topology, uncovering the effects that different knot families have on this collapse. 
We focus on two knot families: double-helix torus knots and twist knots, of increasing complexity~\cite{grosberg_flory-type_1996,stasiak_ideal_1998}, up to 19 crossings, and compare their behavior with that of the unknot. We find that the knot family has a strong impact on the collapse, that becomes more evident with increasing knot complexity, to the point that sufficiently complex twist knots do not support an extended phase, while torus knots do. 
We rationalize this behavior by analyzing the properties of the extended conformations, quantifying the probability of collisions between beads that can lead to deadlocks, and by comparing the extended conformations to those of ideal knots. 

%trace the reason for this behavior back to the stretched configurations of the active knots, with regular aligned strands in the case of torus knots and noose-like constraints for twist knots. 
%%% A STO PUNTO TANTO VALE TOGLIERE LA FRASE
The manuscript is organized as follows: in Sec.~\nameref{sec:model} we go over the details of the model and the simulations, in Sec.~\nameref{sec:results} we present our results regarding the phenomenology of collapse (Sec.~\nameref{ssec:collapse}) and its causes (Sec.~\nameref{ssec:collisions}) and finally in Sec.~\nameref{sec:discussion} we summarize and discuss our conclusions.\\

\section{Methods}
\label{sec:model}
We simulate ring polymers in bulk as closed Kremer-Grest chains of $N$ beads~\cite{grest_molecular_1986} in good solvent, whose topology (i.e. their knot) is set at the beginning of the run. We employ the simulation code LAMMPS~\cite{thompson_lammps_2022}, with a in-house modification for implementing the tangential activity. The beads have diameter $\sigma$, mass $m$, are in contact with a thermal bath with energy $k_BT$ and follow Langevin dynamics with damping time $\tau_\gamma=.3\tau$, where $\tau=\sigma \sqrt{m/k_BT}$ is the unit of time of the system. They interact with each other via a WCA potential
\begin{equation}
    V_{\text{WCA}}\equiv \begin{cases}
    4\epsilon \left[\left(\sigma/r\right)^{12}-\left(\sigma/r\right)^6\right]+\epsilon & r\leq r_{c},\\
    0 & r>r_{c},
\end{cases}
\end{equation}
where $r$ is the distance between monomers, $r_c=\sqrt[6]{2}\sigma$ and $\epsilon=300k_BT$.  Within the polymers, consecutive beads attract each other with a FENE attractive potential:
\begin{equation}
    V_{\text{FENE}}\equiv \begin{cases}-\frac{1}{2}KR_0^2\ln \left[1-\left(r/R_0\right)^2\right] & r\leq R_0,\\
    +\infty & r>R_0,
\end{cases}
\end{equation}
where $K=30\epsilon/\sigma^2$ is the stiffness of the spring and $R_0=1.05\sigma$ is the maximum bond length. The large value of $\epsilon$ and small value of $R_0$ were chosen to stiffen this FENE potential and hence prevent polymers from changing their topology~\cite{locatelli_activity-induced_2021}. Each bead $i$ is furthermore subject to a bending potential
\begin{equation}
V_{b,i}\equiv \kappa(1+\textbf{t}_{i-1}\cdot\textbf{t}_{i}),    
\end{equation}
where $\kappa=k_BT$ is the bending energy and $\textbf{t}_i\equiv (\textbf{r}_{i+1}-\textbf{r}_i)/(|\textbf{r}_{i+1}-\textbf{r}_i|)$ is the normalized tangent vector between neighboring monomers with position $\textbf{r}_i$. Finally, activity is introduced in the system as a constant force pushing each bead $i$ tangentially to the backbone of the polymer~\cite{isele2015self,anand2018structure,janzen2025active} 
\begin{equation}
\textbf{F}_{a,i}\equiv\text{Pe}\frac{ k_BT}{\sigma}(\textbf{t}_{i-1}+\textbf{t}_{i}),    
\end{equation}
where the adimensional { Péclet  number} $\text{Pe}$ quantifies the strength of activity in the system: it was set to $\text{Pe}=0$ for passive systems and $\text{Pe}=10$ for active ones. 

The simulations have a time-step of $\text{d}t=10^{-3}\tau$ and run for an overall time of $10^6 \tau$. As large active molecules tend to collapse and enter a glassy behavior, it becomes unpractical to improve the statistics by increasing the simulation times, as autocorrelation times diverge. To obviate this issue, we simulate instead multiple independent realizations of each molecule, 30 to be specific. Simulations of active polymers start from previously equilibrated passive configurations.

We consider polymers with $20\leq N \leq 1024$ and multiple different knot topologies: twist knots and torus knots.
%\textit{double-helix}\cite{olsen_principle_2013} torus knots ($\mathcal{T}_2$), twist knots and unknots ($0_1$).
Twist knots can be made by repeatedly twisting an unknot around its center and finally clamping the two extremities together~\cite{tubiana_topology_2024}.
Torus knots can be constructed by embedding a curve on a toroidal surface~\cite{oberti_torus_2016}, and are completely defined by the number of windings $h$ around the torus axis of rotational symmetry (the \textit{helices}) and the number of windings $w$ around the torus itself, leading to the naming convention $\mathcal{T}_h^w$. In this manuscript we focus in particular on \textit{double-helix}\cite{olsen_principle_2013} torus knots, $h=2$, reporting also the behavior of two $h=3$ topologies for reference (see SI). The characterization of $T_3^w$ torus knots will however require a second, larger study. 

Knot complexity is loosely captured by the minimal number of crossings, $MCN$, i.e. the minimum number of crossings with which a knot can be drawn on flat surface. For torus knots this follows the relation $MCN=w(h-1)$. %For $h=1$, this yields $MCN=0$, and indeed the unknot can be seen as a single-helix torus knot $\mathcal{T}_1$. 
While $MCN$ provides an intuitive measure of knot complexity, and works intra-family, the number of knots sharing the same value of $MCN$ grows exponentially with it. For this reason in the following we also use the parameter $p=\frac{L}{d}$, the ratio between the length and thickness of an \textit{ideal} knot of a given topology~\cite{grosberg_flory-type_1996,stasiak_ideal_1998}. This parameter has the further advantage of connecting topological and geometrical properties, as it can be interpreted as the minimum length of rope needed to tie a specific knot.

We label all knots with $MCN\leq10$ using the common Alexander-Briggs (\textit{A-B}) notation, where the main number indicates the $MCN$ and the index differentiates between knots with the same $MCN$. For torus knots, this yields $\mathcal{T}_2^w\equiv w_1$ (for example, $\mathcal{T}_2^9\equiv9_1$). High complexity torus knots with $MCN>10$ are not named in the \textit{A-B} frame, so we refer to them with the $\mathcal{T}_2^w$ format. Although the trefoil knot $\mathcal{T}_2^3\equiv3_1$ is a member of both the torus family and the twist family, we decide to group it mainly with the former, as its active behavior has more in common with torus knots. For what regards chiral knots, we used both the left-handed and right-handed chiralities interchangeably, as we observed no relevant difference between the two versions.

\section{Results}
\label{sec:results}
\subsection{The collapse transition}
\label{ssec:collapse}
\begin{figure}
\centering
\includegraphics[width=1.0\textwidth]{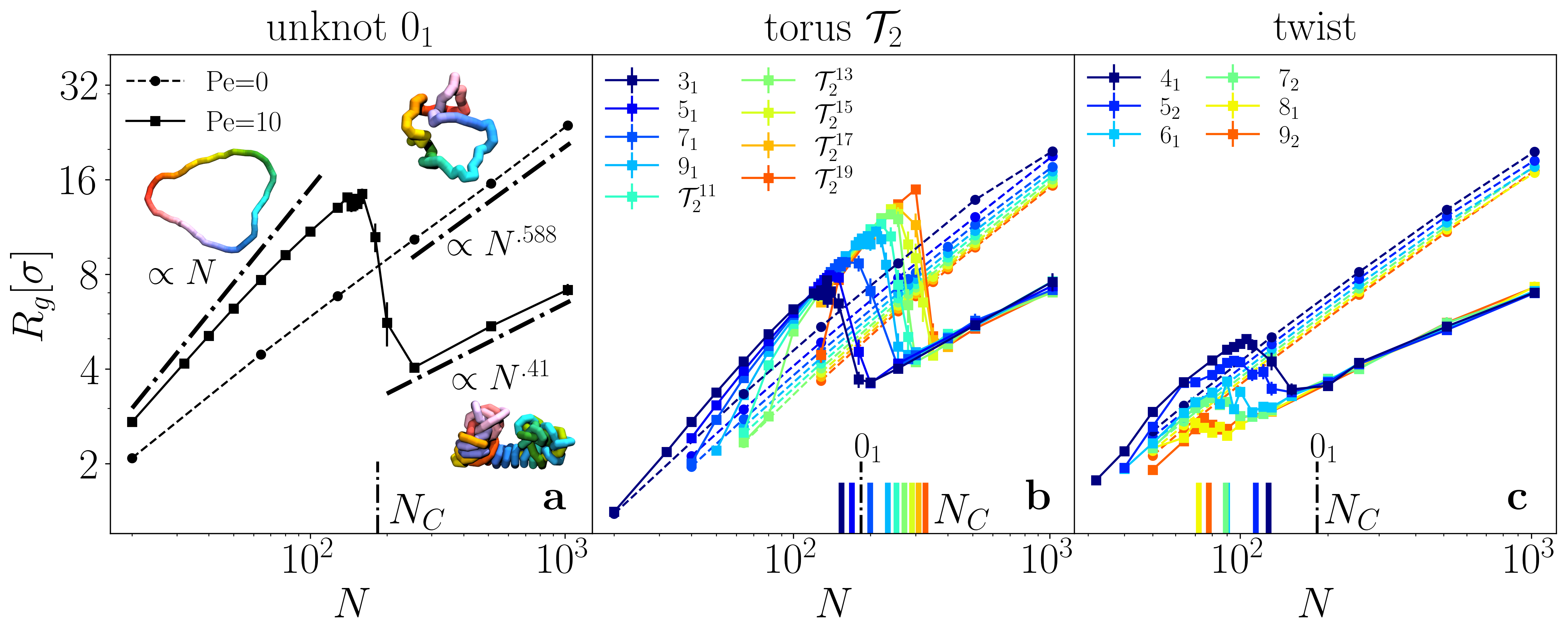}
\caption{Gyration radius $R_g$ of active (---) and passive (- -) ring polymers with various topologies: unknot $0_1$ (a) double-helix torus $\mathcal{T}_2$ (b), and twist (c) knots as a function of polymerization $N$. The lines at the bottom represent the position of the collapsing point $N_C$, with the black longer dash-dot line representing $N_C$ for unknots  (estimation details in SI). The insets in (a) are simulation snapshots of unknots with different configurations: actively stretched ($\text{Pe}=10$, $N=64$, upper left corner), actively collapsed ($\text{Pe}=10$, $N=256$, lower right corner) and passive ($\text{Pe}=0$, $N=64$, upper right corner).}
\label{fig:Rg}
\end{figure}

We quantify the collapse of the polymers by measuring their gyration radius 
\begin{equation*}
R_g\equiv\sqrt{\langle\sum_{i=1}^N(\textbf{r}_i-\textbf{r}_{cm})^2\rangle/N} \, ,    
\end{equation*}
where $\textbf{r}_{cm}$ is the position of the center of mass of the polymer and $\langle\cdot\rangle$ represents the sample average over time and different molecules. We notice in Fig.~\ref{fig:Rg} that knotted polymers show the same qualitative behaviors as unknotted ones: in the passive case there is a single regime, $R_g\propto N^\nu$, with $\nu=0.588$ for our good solvent, while in the active one two different regimes emerge with increasing $N$. First, one finds a stretched regime $R_g\propto N^a$, with $a>\nu$, while a collapsed regime $R_g\propto N^{0.41}$ appears at large enough values of $N$~\cite{locatelli_activity-induced_2021}; the two regimes are separated by a collapse transition at $N_C$.  Looking at Fig.~\ref{fig:Rg}b,c it becomes immediately apparent that this collapse transition happens at a value of $N_C$ that is different for each knot, with torus knots (Fig.~\ref{fig:Rg}b) being more resistant to collapse than twist knots (Fig.~\ref{fig:Rg}c), and with this difference steadily increasing with complexity (i.e. as the $MCN$ grows larger). At small enough values of $N$, the gyration radius of passive and active knots becomes comparable: this is due to the fact that, for each knot, there exists a minimum number of beads with which it can be tied, irrespectively of activity. This number is model-dependent but can be approximated by the smallest integer larger than the length/diameter ratio $p$ of \textit{ideal} knots~\cite{katritch_geometry_1996,stasiak_ideal_1998,olsen_principle_2013}. It was found by Grosberg et al.~\cite{grosberg_flory-type_1996} that this ratio has a determining effect on the gyration radius of passive polymers. In fact, in the scaling law $R_g(N,p)\propto N^\nu p^{-\nu+1/3}$ (see SI), topology only appears as $p$. A similarly straightforward treatment for active polymers is not possible, and an in-depth understanding of the configurations of active polymers of different families becomes necessary to address the behavior of $R_g$ and active collapse.

\begin{figure}
\centering
\includegraphics[width=1.0\textwidth]{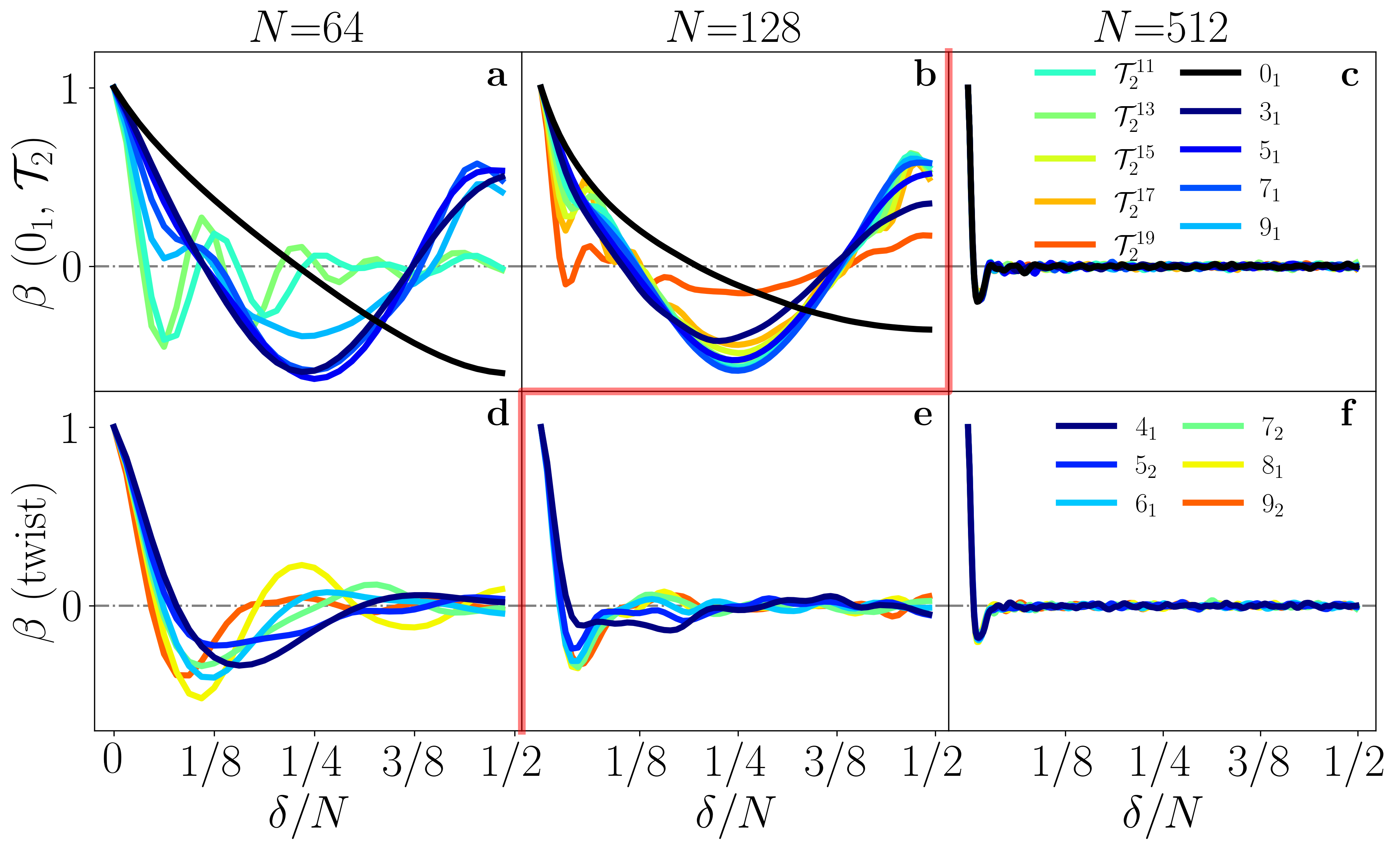}
\caption{Bond correlation function $\beta(\delta)$ of active ring polymers with various topologies: unknot $0_1$, torus $\mathcal{T}_2$ (a,b,c) and twist (d,e,f) for different values of $N$. The red line remarks the collapse transition, which happens before $135\lesssim N_T\lesssim 145$ for all twist knots and after $N_T$ for all $\mathcal{T}_2$ ones. Within this phenomenology, trefoils and unknots behave as torus knots.}
\label{fig:beta1D}
\end{figure}

The effects of activity on polymer configuration, and especially their collapse, become apparent when studying the bond correlation function $\beta(\delta)\equiv \langle \sum_{(i,j)}\textbf{t}_i\cdot  \textbf{t}_j/N\rangle$, where $\delta=|i-j|$ (see Fig.~\ref{fig:beta1D}).  In the first regime, activity induces molecules to rotate at high speeds, as shown by a particularly large angular momentum density (see SI); this is accompanied by an effective increase in bending rigidity and long-distance correlation (Fig.~\ref{fig:beta1D}a,b,d, see SI for the passive case) which lead to stretched polymers and a larger $R_g$.  In the second regime instead, activity causes the total collapse of the molecules into twisted blobs (especially evident in the large torsional parameter - see SI) whose bond orientation decorrelates after $\simeq5$ beads (Fig.~\ref{fig:beta1D}c,e,f), reducing $R_g$ significantly. The collapse transition has furthermore a strong effect on the overall symmetry of the polymers, as stretched active polymers tend to be more oblate (disc-like) than collapsed polymers, with torus polymers being particularly asymmetric (see SI for a discussion on the asphericity and prolateness of the configurations). As already evident from Fig~\ref{fig:Rg}, topology becomes effectively irrelevant in the collapsed state, as $R_g$ falls on the same master curve for all knots; crucially, we note that the collapse transition (delimited in Fig.~\ref{fig:beta1D} by a red line) is strongly dependent on the specific topology of the polymer. In fact, all knots of the torus family (and the unknot) feature the transition at values greater than a certain threshold $135\lesssim N_T\lesssim 145$, while the twist knots collapse at lower values of $N$.\\

\begin{figure}
\centering
\includegraphics[width=0.8\textwidth]{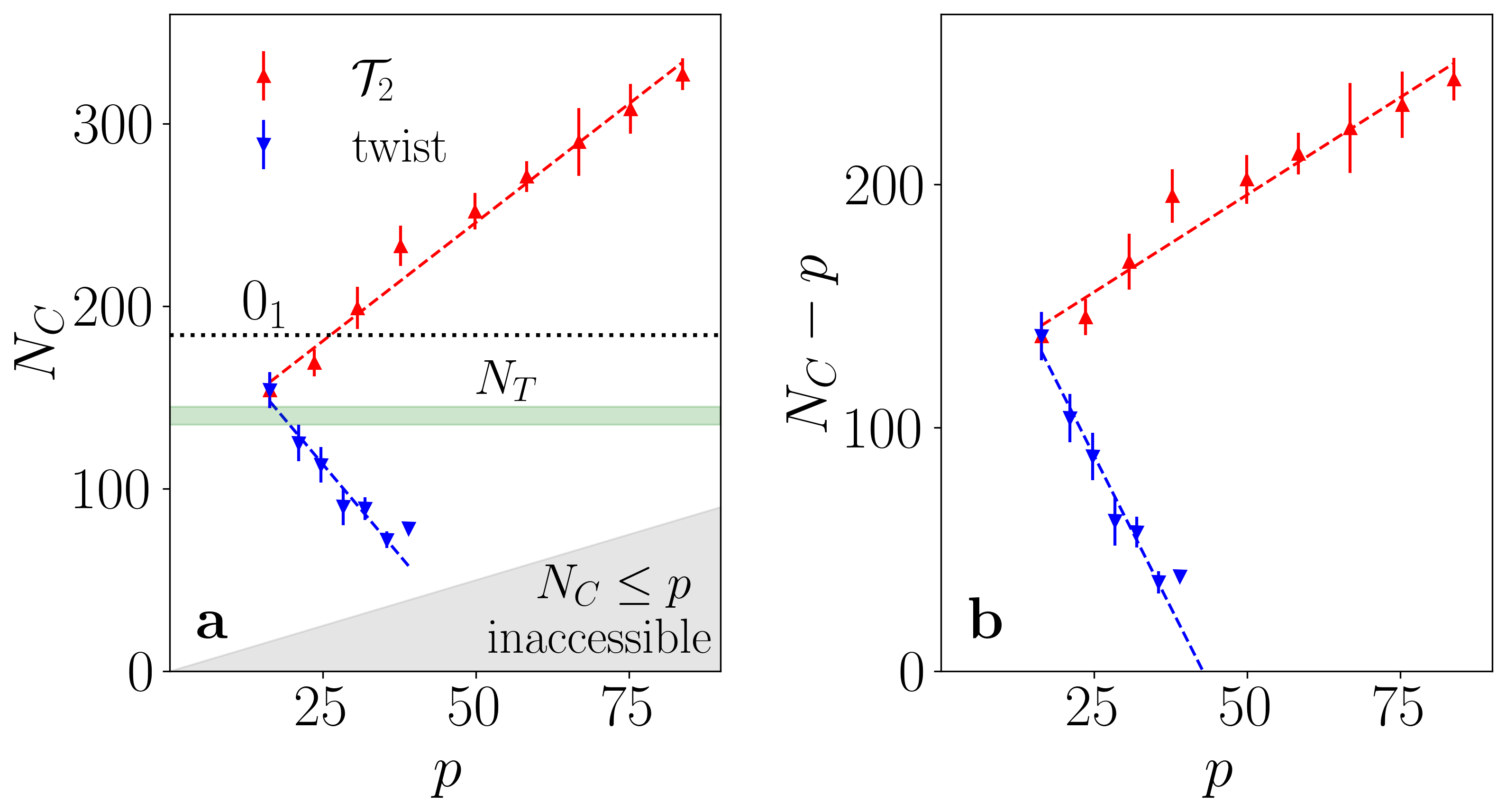}
\caption{(a) Collapsing point $N_C$ for double-helix torus (red upper-facing triangles) and twist knots (blue downward-facing triangles) as a function of the ideal length/diameter ratio $p$. The trefoil knot is shown with both torus and twist symbols. The dotted black line represents the collapsing threshold for the unknot, while the dashed lines are meant to underline the behavior of $N_C$ for the different families. The green area highlights that the collapse of all $\mathcal{T}_2$ knots happens above $N_T$, while that of all twist knots (except the trefoil) takes place below it. The gray area is inaccessible, as all knots must have a number of beads larger than $p$. See SI for the calculation of $N_C$ and its uncertainty. (b) Difference between the collapsing point $N_C$ and $p$ for torus and twist knots as a function of $p$. This difference determines the interval of polymer lengths at which the actively stretched state can exist: for torus knots this regime grows indefinitely, while for twist knots it shrinks with $p$, eventually becoming difficult to measure. The dashed lines are linear fits of the respective families, where the last twist knot ($9_2$) was ignored because of the difficulty of measuring its $N_C$.}
\label{fig:thr_smp}
\end{figure}

Plotting the values of the collapse transition point $N_C$ as a function of their corresponding ideal knot length/diameter ratio $p$  (Fig.~\ref{fig:thr_smp}a, see SI for details on the estimation of $N_C$) highlights the qualitative difference between torus and twist knots, as $N_C$ grows proportionally with $p$ for torus knots, while it decreases for twist knots. This decrease is particularly significant, as it eventually leads to a complete disappearance of the actively stretched regime in twist knots. This phenomenon becomes more evident when one plots the excess length with respect to the length  of the corresponding ideal knot, $(Nc -p)\sigma$ (Fig.~\ref{fig:thr_smp}b). In doing so one can observe that the excess length decreases with $p$ for twist knots, meaning that sufficiently complex twist knots can not maintain extended configurations,  but directly  collapse from their ideal configuration. This can be already seen for knots $8_1$ and $9_2$: their gyration radius growth in the stretched state appears as just a small bump (Fig.~\ref{fig:Rg}c), and we expect this regime to finally disappear for $p\gtrsim 40$, roughly corresponding to $MCN\gtrsim 10$. On the other hand, for torus knots $N_C$ grows faster than $p$ (Fig.~\ref{fig:thr_smp}b), yielding a consistent stretched regime even at high complexity.\\
\subsection{Bead collisions and loop formation}
\label{ssec:collisions}
To better understand the reason behind the difference in behavior between torus and twist knots, we measure how prone different systems are to bead-bead collisions. Collisions in colloidal active systems are known to trigger the formation of clusters at sufficiently high values of the  Péclet number, in a process called Motility-induced Phase Separation (MIPS)~\cite{buttinoni_dynamical_2013,cates_motility-induced_2015}.  In melts of active rings, collisions of strands moving in different directions can also originate \textit{deadlocks}~\cite{micheletti_topology-based_2024}, which prevent the molecules from relaxing and foster additional entanglements. Deadlocks in melt of rings are caused by one ring threading into another one, while in our case there is a just single strand passing inside a loop formed by another portion of the same ring.  Still, a similar principle applies: due to activity this loop may get smaller and smaller, if the local conformation is strongly perturbed, thus stopping the motion of the strand, as it happens for unknotted rings~\cite{locatelli_activity-induced_2021}. At the same time, the loop can't be resolved, as it is topologically prevented by the strand passing through it. 
The stability of the resulting collapsed state could result from the effective bending caused by activity, in a mechanism similar to that of passive tight knots in semiflexible polymers~\cite{de_gennes_tight_1984,grosberg_metastable_2007,dai_metastable_2014}.  

\begin{figure}
\centering
\includegraphics[width=1.0\textwidth]{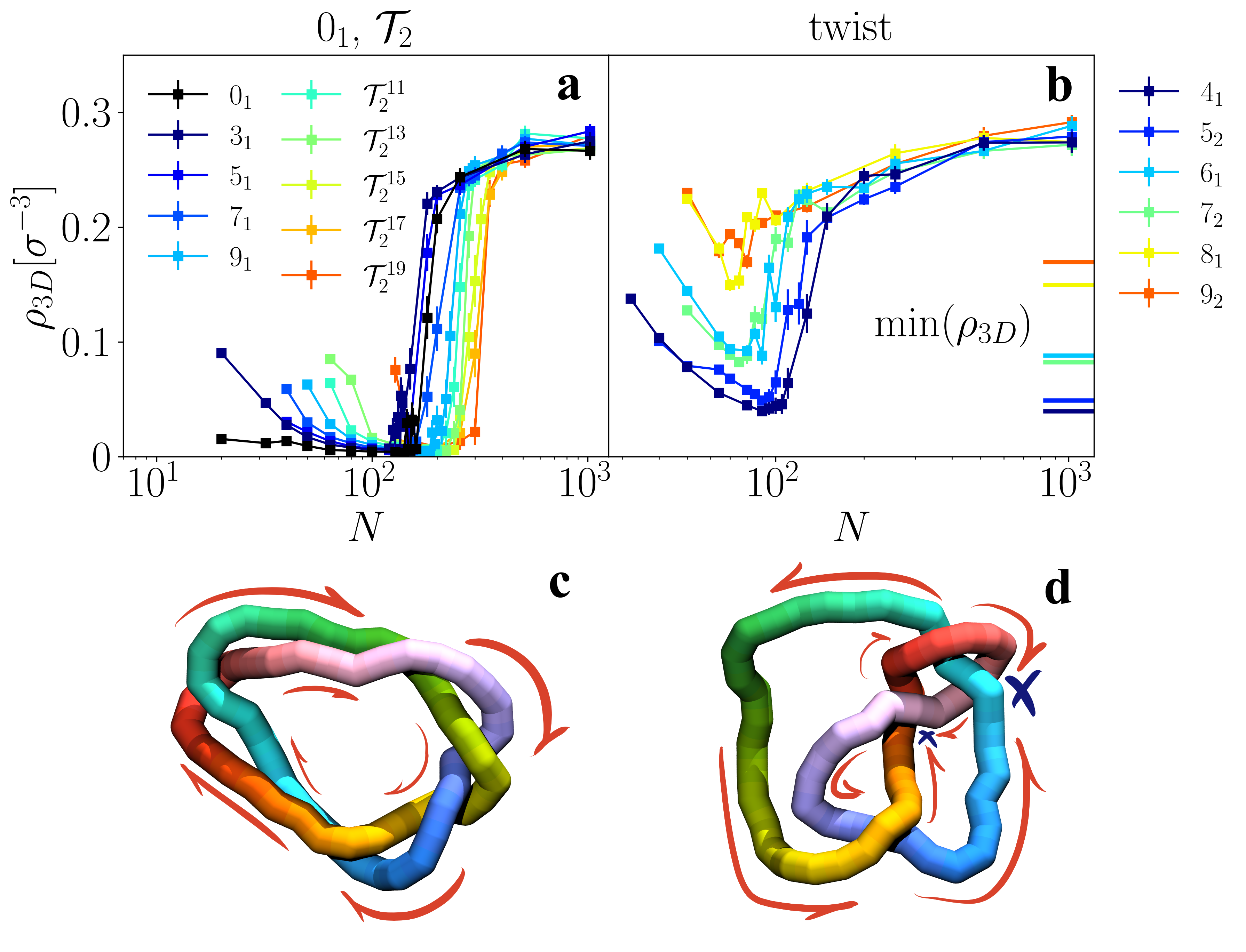}
\caption{(a,b) Number density of close bonds oppositely oriented $\rho_{\text{3D}}$ of active ring polymers with various topologies: unknot $0_1$, torus  $\mathcal{T}_2$ (a), and twist (b) knots as functions of $N$. All torus knots (and the unknot) present a vanishing probability of colliding strands in their stretched state, which is not the case for twist knots. The horizontal lines in (b) indicate the minimum value of $\rho_{3D}(N)$ for different twist topologies, showing that larger complexity twist knots have higher likelihood of colliding strands. (c,d) snapshots of a $5_1$ torus knot (c) and a $4_1$ twist knot (d) in their actively stretched regime, with $N=64$ and $\text{Pe}=10$. The red lines sketch the direction of active motion, while the blue crosses represent points of collision. The regular conformation of the torus eases friction, while the twist knot presents multiple points of collisions, especially nearby its noose-like constraint.}
\label{fig:betarho}
\end{figure}

We measure the probability of collisions by calculating the number density of close bonds (in 3D space) oppositely oriented  $\rho_{\text{3D}}\equiv \langle (\#\{(i,j)\in P_b:\textbf{t}_i\cdot  \textbf{t}_j<0\})/V_P\rangle$, where $V_P=N\pi4\sigma^3$ is the volume of a tube of radius $2\sigma$ and length $N\sigma$. We only consider the set $P_b$ of  non-neighboring bonds ($|i-j|>3$) within a distance in 3D space of $2\sigma$ from each other ($|(\textbf{r}_i+\textbf{r}_{i+1})/2-(\textbf{r}_j+\textbf{r}_{j+1})/2|<2\sigma$). This pair selection $P_b$ is designed to measure the correlation of bonds $\textbf{t}_i \cdot \textbf{t}_j$ in different sections of the polymer that come in contact with each other. 
In fact, as beads actively move in the direction of the polymer backbone, this orientational correlation contains information on intra-molecular collisions, and more specifically tells us if two strands close to each other are likely to collide (null or negative correlation) or are moving in the same direction (positive correlation). Fig.~\ref{fig:betarho} shows how $\rho_{3D}$ behaves as a function of $N$. In all cases $\rho_{3D}$ initially decreases as the polymer extends further, reaches a minimum just before collapse and finally tends towards a value of around $\rho_{3D}=0.3\sigma^{-3}$. The difference between families can be found in the value that this number density takes before collapse: torus and unknots reach smaller values than twist knots, and their strands have a consequently lower likelihood of colliding. Indeed, the minimum of $\rho_{3D}$ is $\approx 0$ for all torus knots considered, regardless of $p$. In contrast, in twist knots the likelihood of collisions increases with $p$ (and therefore with $MCN$), as can be seen in Fig.~\ref{fig:betarho}b.\\

While the behavior of $\rho_{3D}$ sheds light on the mechanical reasons of the collapse, it still does not explain why torus knots are much less susceptible to collisions than twist knots, and also why the torus $N_C$ grows linearly with $p$. The answer to these questions lies in the configuration of the stretched polymers, and especially their bond correlation $\beta$ (see Fig.~\ref{fig:beta1D}). In particular, the point where the bond correlation $\beta$ reaches its minimum, or argmin$(\beta)$, determines the typical loop size of the polymer $N_{l}=2\langle\text{argmin}(\beta)\rangle$. We notice in Fig.~\ref{fig:beta1D}b that for the unknot $\beta$ reaches a minimum around $\delta/N=1/2$, yielding a loop length $N_l$ which almost coincides with $N$, as the full polymer turns into a loop (Fig.~\ref{fig:Rg}a, top left). For $\mathcal{T}_2$ knots the minimum is instead reached at $\delta/N=1/4$, compatible with a configuration resembling that of two loops intertwined in a double-helical shape (Fig.~\ref{fig:betarho}c). This very regular shape, where strands moving actively in the same direction are intertwined, reduces $\rho_{3D}$ to a minimum and makes the polymers less prone to collapsing. 

When $N$ grows, the pitch of the double helix grows as well, and with it the length of loops and the probability that some of them collide, collapsing the polymer. Increasing $p$ lowers the helix's pitch, and therefore the frequency of collisions. As a result, $N_C$ increases on average by $\simeq 20$ beads with each pair of additional crossings. This length of 20 beads is particularly interesting, as it is the same one individuated by Locatelli et al. as the typical $\text{argmin}(\beta)$ at which active ring polymers starting in a passive state either extend further or collapse~\cite{locatelli_activity-induced_2021}
%For low values of $p$, the topological constraints are not enough, with increasing $N$, to keep the polymers in this configuration; however, as the knot complexity increases, the configuration becomes more regular and $N_C$ grows, increasing on average of $\simeq20$ beads  with each additional crossing. %\db{This length is particularly interesting, as it is the same one individuated by Locatelli et al. as the typical loop length at which active ring polymers starting in a passive state either extend further or collapse~\cite{locatelli_activity-induced_2021}.} 
Twist knots, on the other hand, show less regular configurations, and have no mechanism comparable to the one reducing $\rho_{3D}$ in torus knots. On the contrary, they present noose-like constraints that constitute a systematical source for collisions and deadlocking (Fig.~\ref{fig:betarho}d), with a consequently larger probability  of collapse that further grows with the topological complexity, and hence $p$.

\section{Discussion}
\label{sec:discussion}
In this manuscript we have studied how topology affects the collapsing behavior of tangentially active ring polymers. We found that torus ($\mathcal{T}_2$) and twist knots behave in opposing ways, as the collapsing points $N_C$ of the former family are all larger than those of the latter. Moreover, as the ideal knot length/diameter ratio $p$ of the knots increases, the $N_C$ for torus knots grows linearly, making them more resistant to collapse, while that of twist knots decreases, effectively erasing the stretched regime for $p\gtrsim 40$, or $MCN \gtrsim 10$.
We identified the reason for this phenomenology in the configurations that different topologies take when actively stretched, and in the way these configurations foster or hinder bead collisions. In fact, we showed that torus knots tend to have regular double-helix configurations, with intertwined strands that move in the same direction, while twist knots present configurations where the presence of topological constraints, especially noose-like ones, acts as a catalyst for bead collisions, leading to MIPS-like clustering and deadlocking. In the future it will be important to test these phenomena in presence of hydrodynamic interactions, as they have been shown to affect the configuration of ring polymers~\cite{weiss_hydrodynamics_2019,liebetreu_hydrodynamic_2020}. While this was currently unfeasible simply for the number of simulations required by a systematic characterization of torus and twist knots, we are confident that the qualitative message reported in this work will not change when hydrodynamic is explicitly considered, as the double-helix conformation of torus knots minimizes their energy and twist knots are topologically forced to have regions increasing the frequency of deadlock.

%Our results expand significantly on those presented by Locatelli et al.\cite{locatelli_activity-induced_2021} on the collapse of active rings, providing a way to control it by tuning the rings' topology~\cite{cardelli_heteropolymer_2018}. In fact, one could decide to produce polymers which collapse at arbitrarily large lengths by using torus topologies with enough crossings, or to completely suppress the stretched behavior by choosing instead a twist topology. Furthermore, given the distinctly different behavior of different knots, another application would be in the detection of different topologies by providing passive ring polymers with activity, by means of molecular motors. The disappearance of extended twist knots also suggests that the knot spectrum could be used to detect the phase of randomly knotted active polymers, as complex twist knots would be exceedingly rare. 

Our results widens the understanding of the interplay between activity and topology. First of all, they provide a way control the extended regime of active polymers by tuning their topology\cite{cardelli_heteropolymer_2018}. In fact, one could decide to produce polymers which collapse at arbitrarily large lengths by using torus topologies with enough crossings, or to completely suppress the stretched behavior by choosing instead a twist topology. %Furthermore, given the distinctly different behavior of different knots, another application would be in the detection of different topologies by providing passive ring polymers with activity, by means of molecular motors. 
The disappearance of extended twist knots also suggests that the knot spectrum could be used to detect the phase of randomly knotted active polymers, as complex twist knots would be exceedingly rare, or that activity could be used to select only torus knots. 

Finally, this work could be extended to specific active materials where the activity and topology are both important. For example, it would be interesting 
to consider passive-active ring diblock polymers, testing the stability of active ring polymers as the percentage and distribution of active elements changes~\cite{kumar_local_2023}. This would also add on the picture given by the multiple studies on the formation of knots in linear diblock copolymers~\cite{shen_knotting_2025,vatin_upsurge_2025}. Notably, molecular motors could  effectively be modeled by tangential active beads moving along the backbone, pushing the polymer in the opposite direction. In such a case, knotted topologies would constitute an especially interesting playground, as molecular motors have been shown to also diffuse in 3D space\cite{mirny_how_2009,suter_transcription_2020}, and the topological constraints would interplay and possibly facilitate such transport. %Another possible avenue of research would be to study whether topological friction\cite{patil_topological_2020} has a role in bead collision mechanics and polymer collapse. In fact, topologically-induced friction has been found to affect DNA ejection\cite{marenduzzo_topological_2013} and and pore translocation\cite{suma_pore_2015}, both non-equilibrium phenomena. 

%would like to stress that a thorough study of a large amount of topologies such as what we presented in this study would have been difficult without the simplification of an implicit solvent.%Finally, a continuation of this work would be the study of active rings with composite knot topology, or active links. In particular, the study active links could shed light on the active process behind the formation of Olympic gels such as the kDNA~\cite{ramakrishnan_single-molecule_2024}, and open new avenues for the production of active linked supramolecules which could tune their properties as their environment change.\\
%\tableofcontents

%\nocite{*}

\begin{acknowledgement}

This work has been supported by the project “SCOPE—Selective Capture Of Metals by Polymeric spongEs” funded by the MIUR Progetti di Ricerca di Rilevante Interesse Nazionale (PRIN) Bando 2022 (Grant No. 2022RYP9YT). The authors acknowledge the CINECA award under the ISCRA initiative, for the availability of high-performance computing resources and support. D.B. and L.T. thank the UniTN HPC cluster for offering the computing resources. L.T. acknowledges support from ICSC – Centro Nazionale di Ricerca in HPC, Big Data and Quantum computing, funded by the European Union under NextGenerationEU.
\end{acknowledgement}

\begin{suppinfo}
In the Supplemental Information (SI) we explain in detail the calculation of the collapse point $N_C$, we show additional properties of active knots, such as their torsional order parameter $U_T$, angular momentum density $l$, asphericity $A$ and prolateness $P$. Furthermore, we show the gyration radius $R_g$ and bond correlation function $\beta$ for passive knots and the behavior of the first two triple-helix $\mathcal{T}_3$ knots.

\end{suppinfo}

\bibliography{acknots}

\newpage

\title{}
%%%%
\clearpage
\begin{center}
\textbf{\Large  \vspace*{1.5mm} Effects of knotting on the collapse of active ring polymers -- Supplemental Information } \\
\vspace*{5mm}
Davide Breoni, Emanuele Locatelli, and Luca Tubiana
%Authors
\vspace*{10mm}
\end{center}
%%%

\subsection{Determination of the collapsing point $N_C$}
\begin{figure}
\centering
\includegraphics[width=1.0\textwidth]{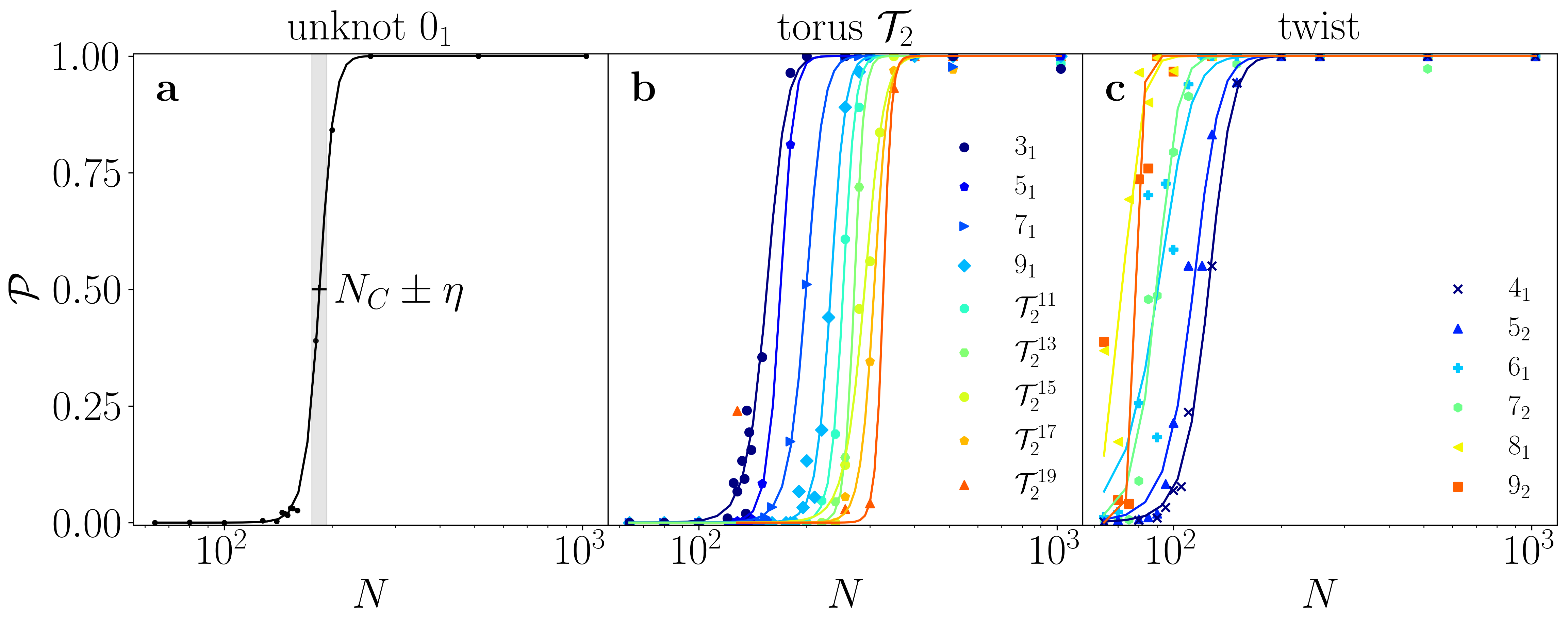}
\caption{Probability $\mathcal{P}\equiv P(R_g<0.9\langle R_g^0\rangle)$ for double-helix torus (a), unknot and triple-helix torus (b), twist knots as a function of $N$. The shaded area in (a) delimits the position and uncertainty of the collapsing point $N_C\pm\eta$. }
\label{fig:Rg_hist}
\end{figure}

To determine the collapse point $N_C$ of an active polymer with a given knot, we use a special ``cumulative'' distribution of $R_g$, that is, the probability that the observed $R_g$ is smaller than the average equilibrium value, $\mathcal{P}(N) \equiv P(R_g<0.9\langle R_g^0(N)\rangle)$. 
We notice in fact that, before collapse, active knots generally feature $\langle R_g\rangle>\langle R_g^0\rangle$, while after the collapse the opposite holds. As such, the probability curve  increases from 0 to 1 as $N$ grows, allowing us to extract $N_C$ by fitting this curve to a sigmoid function $\mathcal{S}(N)\equiv (1 + \text{exp}((N_C-N)/\eta))^{-1}$ (see Fig.~\ref{fig:Rg_hist}).
The sigmoid parameter $\eta$ was used as uncertainty for the estimation of $N_C$. The 0.9 factor in the formula for $\mathcal{P}$ is necessary to keep $\mathcal{P}$ close to 0 for the small values of $N$ at which $R_g$ and $R_g^0$ tend to be very similar.
\subsection{Torsional order parameter $U_T$}
\begin{figure}[h]
\centering
\includegraphics[width=0.8\textwidth]{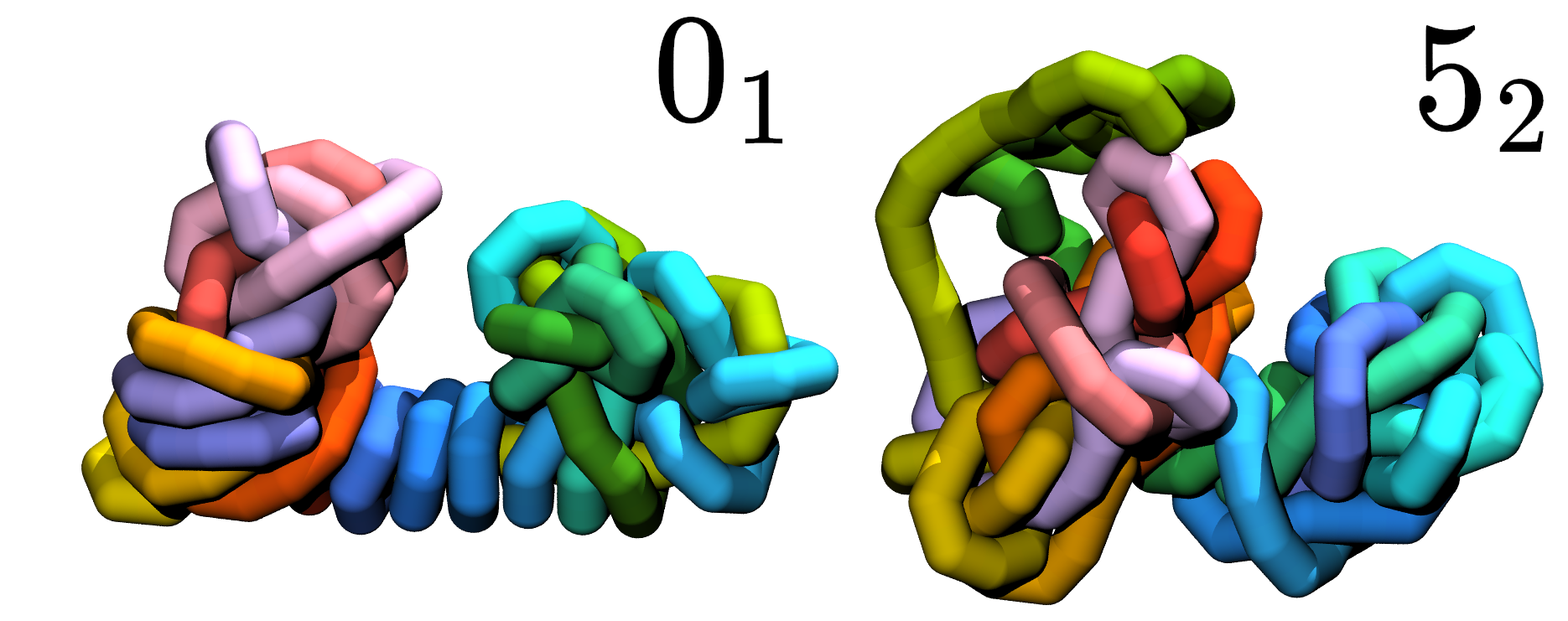}
\caption{Collapsed conformation of active ring polymers with activity $Pe=10$ and polymerization $N=256$: an unknot (left) and a $5_2$ twist knot (right). Color represents the bead index.}
\label{fig:twist}
\end{figure}
\begin{figure}[ht]
\centering
\includegraphics[width=1.0\textwidth]{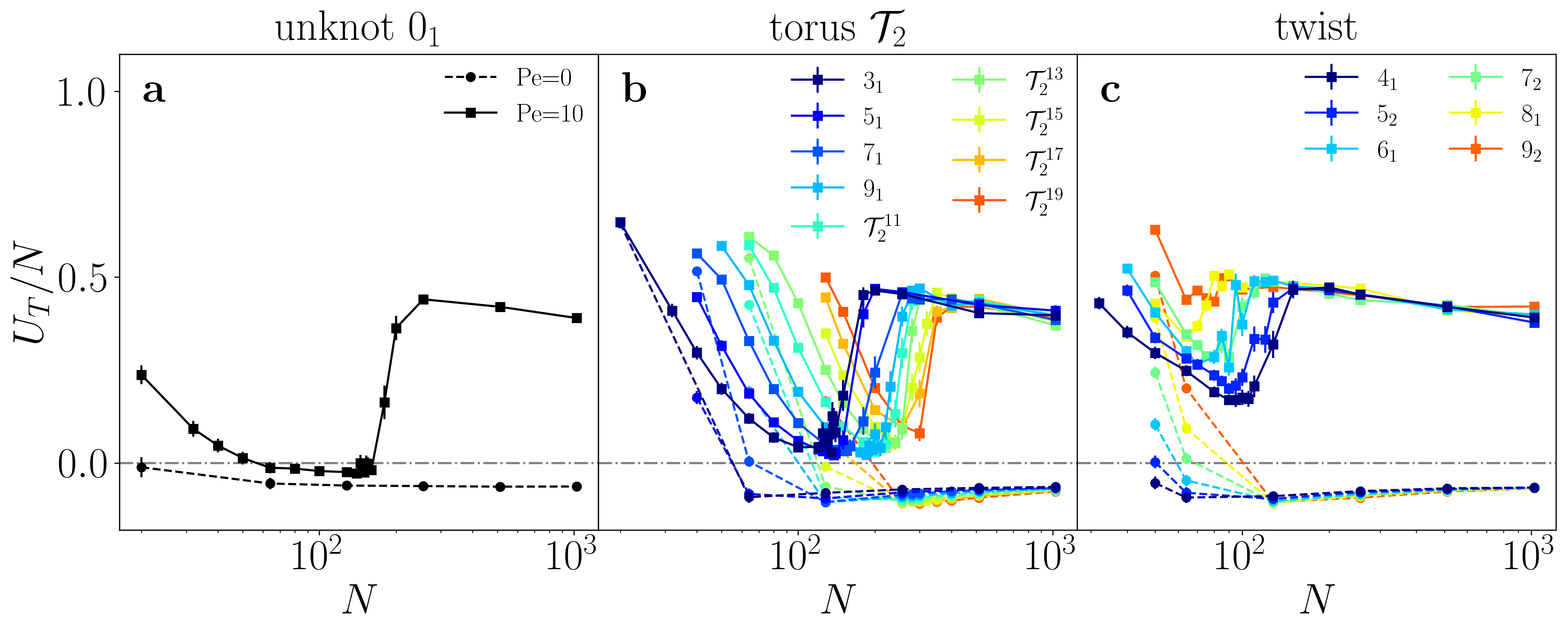}
\caption{Torsional order parameter $U_T/N$ of active (---) and passive (- -) ring polymers with various topologies: unknot $0_1$ (a), double-helix torus $\mathcal{T}_2$ (b), and twist (c) knot as a function of $N$.}
\label{fig:tors}
\end{figure}

In the collapsed state, active polymers fall into a heavily twisted conformation (see Fig.~\ref{fig:twist}). This twisting can be quantified with the torsional order parameter $U_T$:
\begin{equation}
    U_T\equiv\left\langle\sum_i^N\frac{(\textbf{t}_{i-1}\times\textbf{t}_i)\cdot(\textbf{t}_{i}\times\textbf{t}_{i+1})}{|\textbf{t}_{i-1}\times\textbf{t}_i||\textbf{t}_{i}\times\textbf{t}_{i+1}|}\right\rangle,
\end{equation}
where $\textbf{t}_j\equiv (\textbf{r}_{j+1}-\textbf{r}_j)/(|\textbf{r}_{j+1}-\textbf{r}_j|)$ is the normalized tangent vector between neighboring monomers with position $\textbf{r}_j$, $N$ is the number of bead per polymer and $\langle\cdot\rangle$ represents averaging over time and different molecules. We observe that both the ideal and collapsed state are extremely twisted states, while polymers tend to minimize $U_T/N$ in their actively stretched configurations (Fig.~\ref{fig:tors}).

\subsection{Angular momentum density $l$}
\begin{figure}[ht]
\centering
\includegraphics[width=1.0\textwidth]{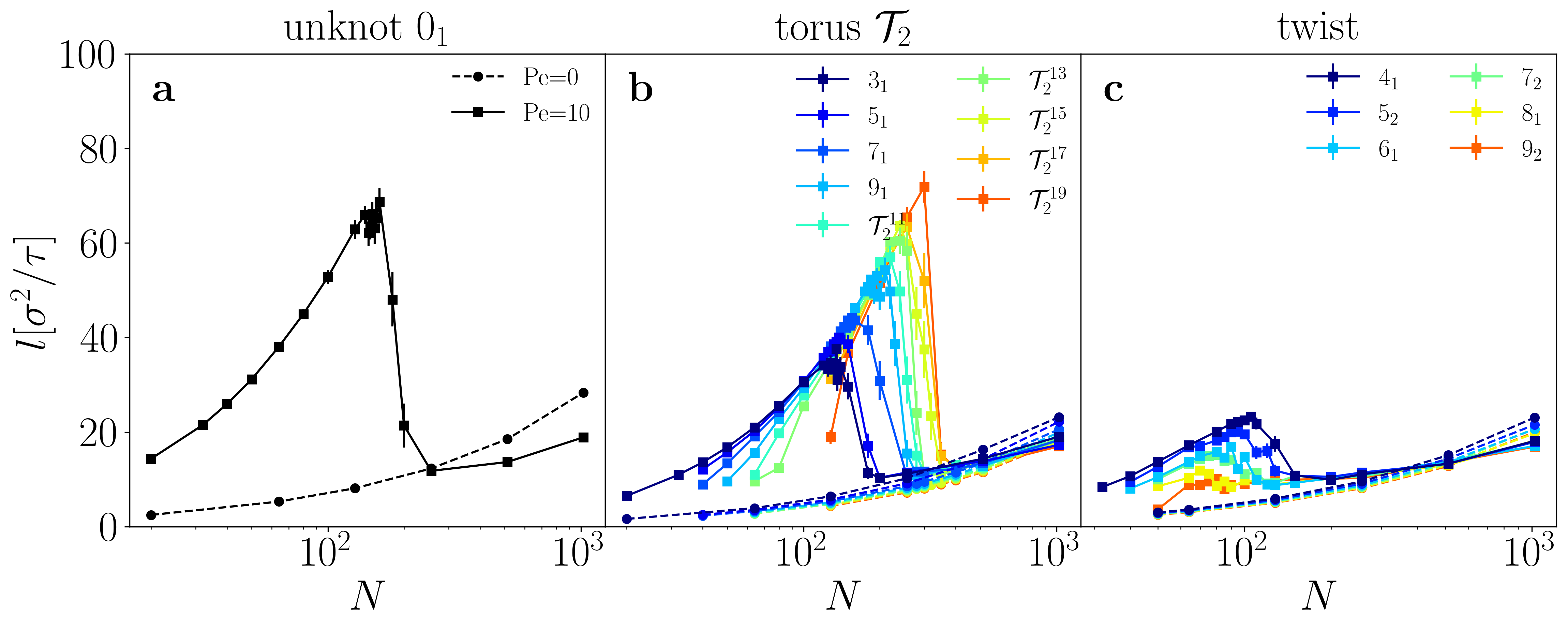}
\caption{Angular momentum density $l$ of active (---) and passive (- -) ring polymers with various topologies: unknot $0_1$ (a), double-helix torus $\mathcal{T}_2$ (b), and twist (c) knot as a function of $N$.}
\label{fig:angmom}
\end{figure}

When in the stretched state, the configuration of the active polymers and the tangential forces allow them to rotate very fast around their normal axis. We measure this with the help of the angular momentum density $l$, defined as:
\begin{equation}
    l\equiv\left\langle\frac{m}{N}\left| \sum_i^N(\textbf{r}_{i}-\textbf{r}_{cm})(\textbf{v}_i-\textbf{v}_{cm})\right|\right\rangle,
\end{equation}
where $m$ is the mass of the beads, $\textbf{v}_i$ is the velocity of bead $i$, and $\textbf{r}_{cm}$ and $\textbf{v}_{cm}$ are respectively the position and velocity of the center of mass of the polymer. We notice that, for active polymers, $l$ increases constantly until the collapse transition. Furthermore, unknots and $\mathcal{T}_2$ knots can reach significantly larger angular momenta than twist knots, as their monomers can rotate further away from the rotation axis (see Fig.~\ref{fig:angmom}).
\subsection{Asphericity $A$ and prolateness $P$}
\begin{figure}[ht]
\centering
\includegraphics[width=.7\textwidth]{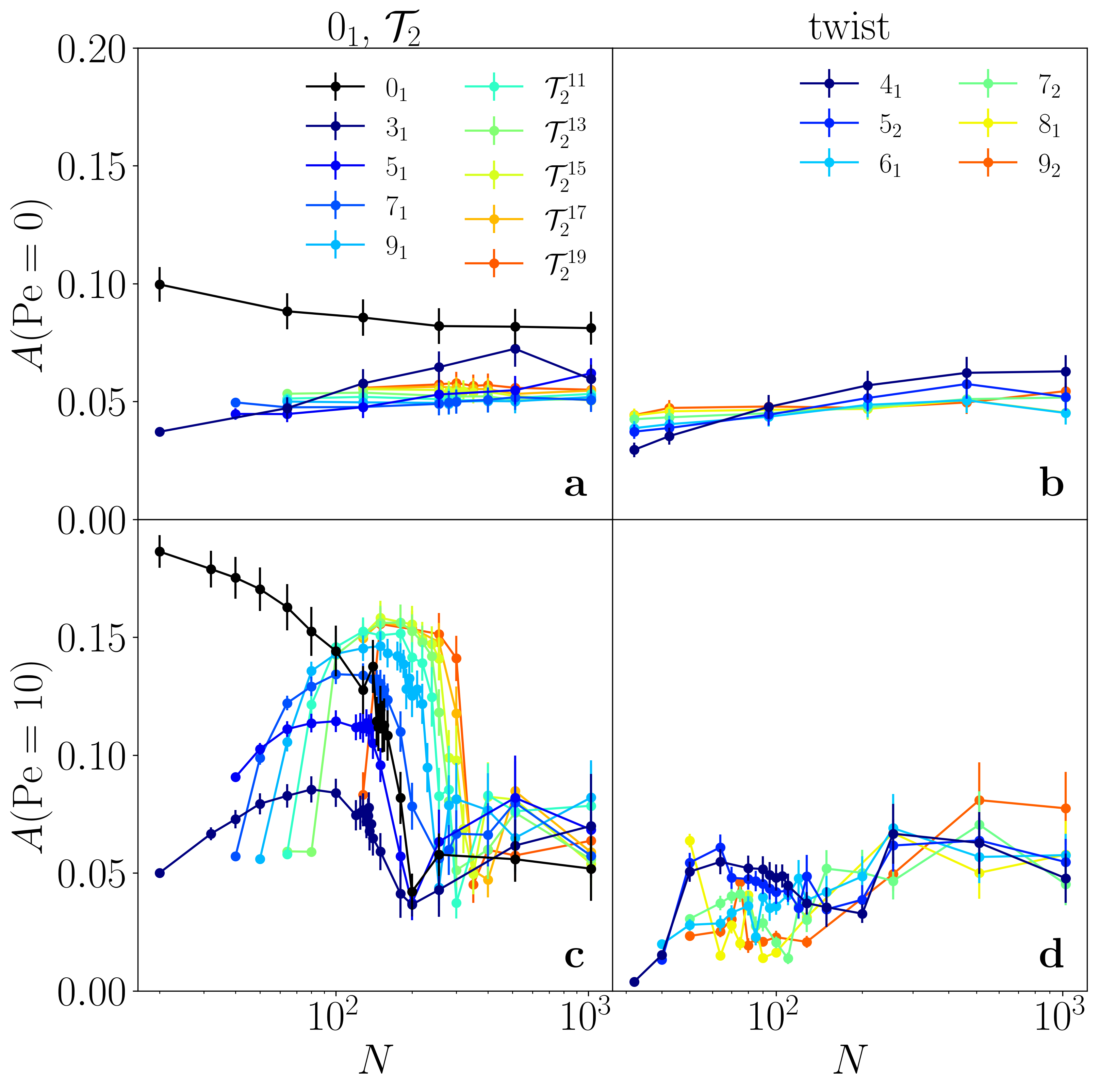}
\caption{Asphericity $A$ of passive (a,b) and active (c,d) ring polymers with various topologies: unknot $0_1$, double-helix torus $\mathcal{T}_2$ (a,c), and twist (b,d) knots as a function of $N$.}
\label{fig:asph}
\end{figure}

It was found by Rawdon et al.~\cite{rawdon_effect_2008,millett_effect_2009} that knot topology has an important effect on the conformation of passive knots, and more specifically the shape of their ellipsoid of inertia. Similarly, we want to measure the effects of activity on the inertial ellipsoid of our polymers, and in order to do so, we calculate two quantities: the asphericity $A$, which gauges how spherically asymmetric the polymer is, and the prolateness $P$, which distinguishes between oblate objects (short and wide) and prolate objects (thin and long). Both quantities are computed starting from the moment of inertia tensor $T_{\alpha\beta}$, defined as
\begin{eqnarray}
\label{Tab}
    T_{\alpha\beta} &\equiv& \frac{1}{N}\left\langle\sum_{i=1}^{N} (r^\alpha_{i}-r^\alpha_{cm})(r^\beta_{i}-r^\beta_{cm})\right \rangle,
\end{eqnarray}
where $\textbf{r}_{cm}$ is the center of mass of the polymer and $\alpha$ and $\beta$ indicate the Cartesian axes. We then calculate the eigenvalues of $T_{\alpha\beta}$, whose square roots define the three semiaxes of the inertial ellipsoid: $a$, $b$ and $c$. The asphericity $A$ is then
\begin{eqnarray}
\label{eq:as}
    A\equiv \frac{(a-b)^2+(b-c)^2+(c-a)^2}{2(a+b+c)^2},
\end{eqnarray}
where $A=0$ defines a spherically symmetric shape and $A=1$ describes instead a rod-like shape. We observe that in their actively stretched state torus knots and unknots (Fig.~\ref{fig:asph}c) are much more aspherical than twist knots (Fig.~\ref{fig:asph}d) and their passive counterparts (Fig.~\ref{fig:asph}a,b). This is a result of the rather flat torus double-helix configurations.

\begin{figure}[ht]
\centering
\includegraphics[width=.7\textwidth]{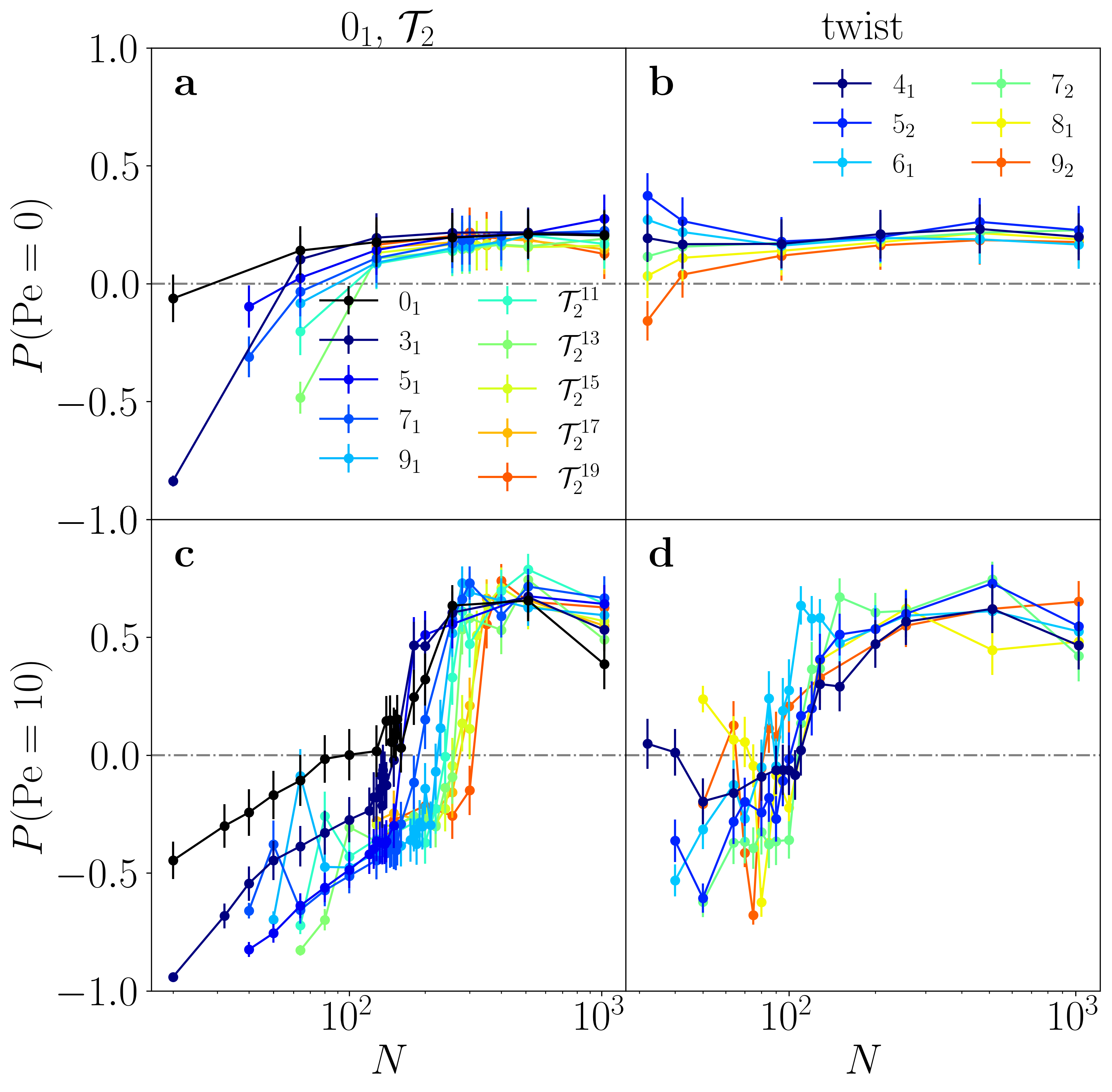}
\caption{Prolateness $P$ of passive (a,b) and active (c,d) ring polymers with various topologies: unknot $0_1$, double-helix torus $\mathcal{T}_2$ (a,c), and twist (b,d) knots as a function of $N$.}
\label{fig:prol}
\end{figure}

The prolateness $P$ is measured as 
\begin{eqnarray}
\label{eq:prol}
    P\equiv \frac{(2a-b-c)(2b-a-c)(2c-a-b)}{2(a^2+b^2+c^2-ab-ac-bc)^{3/2}},
\end{eqnarray}
going from $P=-1$ for an oblate disc-like object to $P=1$ for a prolate rod-like object. We notice that all active knots tend to go from an oblate state to a prolate one as they collapse (Fig.~\ref{fig:prol}c,d), while the same is not always true for passive knots (Fig.~\ref{fig:prol}a,b). Furthermore, active torus stretched configurations tend to be more disc-like than twist configurations.
\subsection{Gyration radius and bond correlation function of passive polymers}
\begin{figure}[ht]
\centering
\includegraphics[width=1.0\textwidth]{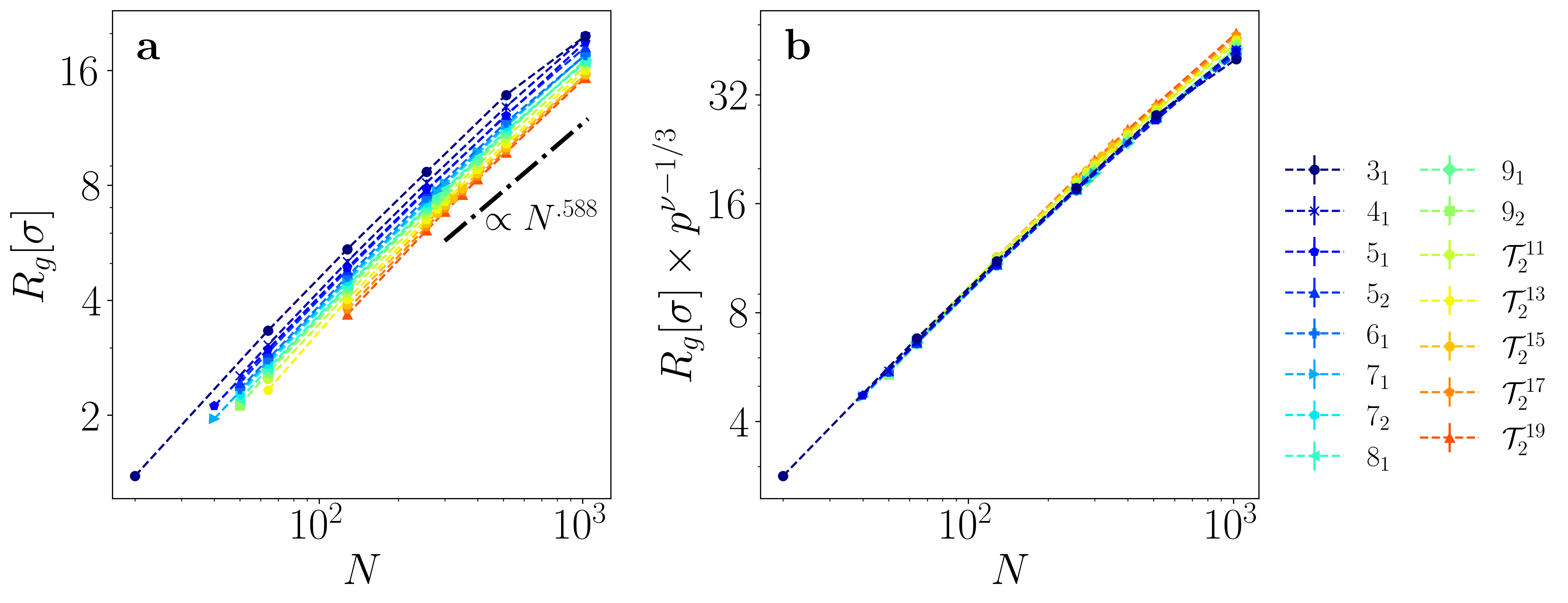}
\caption{$R_g$ of passive ring polymers with various topologies as a function of polymerization $N$ (a) and rescaled to take into account knot complexity (b).}
\label{fig:Rg_passive}
\end{figure}

\begin{figure}
\centering
\includegraphics[width=1.0\textwidth]{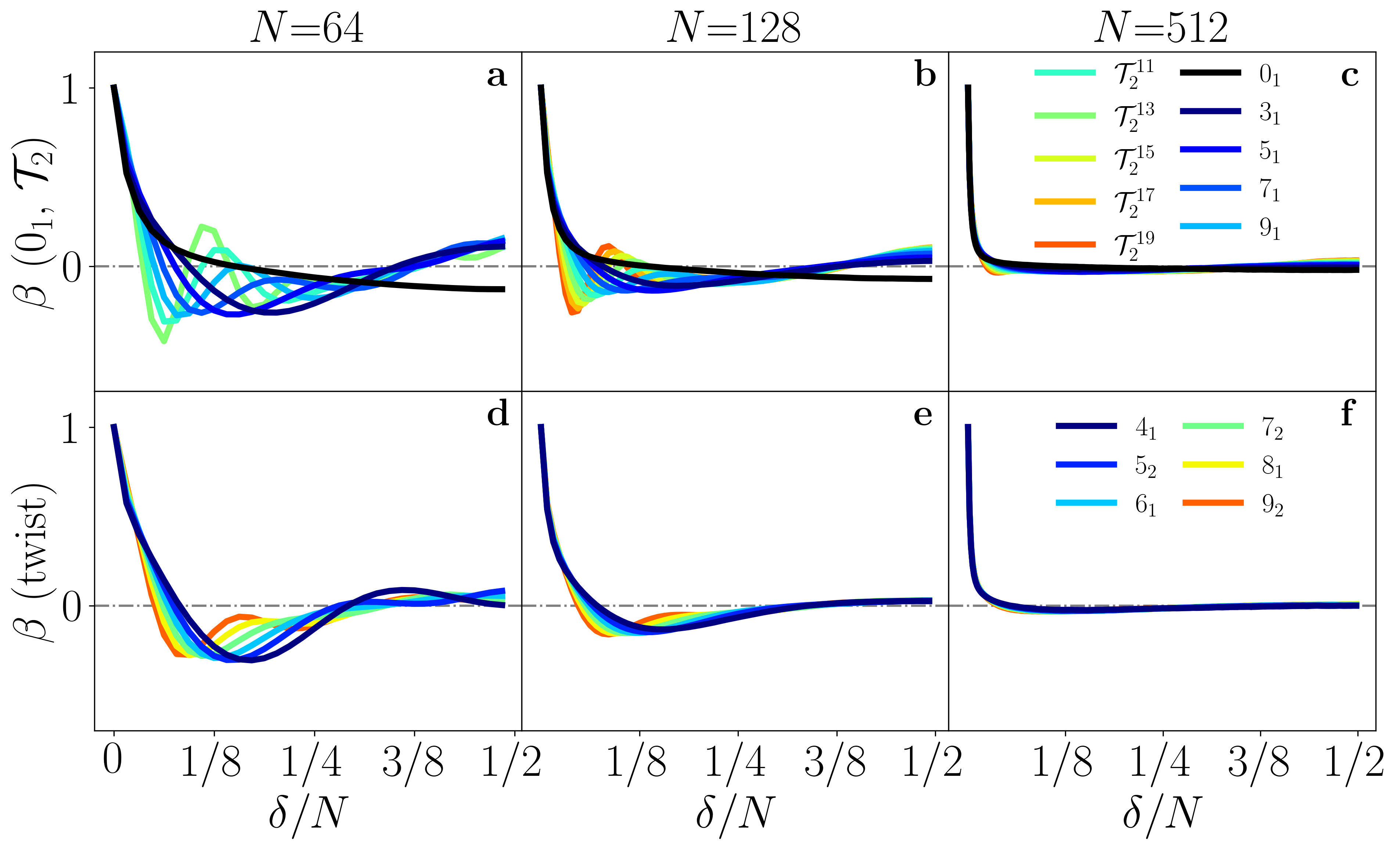}
\caption{Bond correlation function $\beta(\delta)$ of passive ring polymers with various topologies: unknot $0_1$, torus $\mathcal{T}_2$ (a,b,c) and twist (d,e,f) for different values of $N$.}
\label{fig:beta1D_0}
\end{figure}

In Fig.~\ref{fig:Rg} of the main manuscript we see that the gyration radius $R_g$ of passive ring polymers scales with $N^\nu$, where $\nu=0.588$, as typical of self-avoiding ring polymers in good solvent. Furthermore, we notice that $R_g$ depends on the topology of the rings, decreasing as the complexity of the knot increases. This complexity can be quantified by taking the smallest length/diameter ratio $p$ of the knot in its ideal configuration~\cite{katritch_geometry_1996,stasiak_ideal_1998,olsen_principle_2013}. The dependence of $R_g$ as a function of $p$ was theoretically calculated by Grosberg et al. in Ref.~\cite{grosberg_flory-type_1996}, yielding $R_g(N,p)\propto N^\nu p^{-\nu+1/3}$. We confirm this behavior in Fig.~\ref{fig:Rg_passive}b, where $R_g$ is rescaled accordingly, and as a result all knot topologies follow the same master curve.  We exclude from this plot the unknot, for which $p$ is not well defined.\\
To complete the information given by Fig.~\ref{fig:beta1D} on the bond correlation function, we show in Fig.~\ref{fig:beta1D_0} the results for passive systems with the same topologies and number of beads $N$. We notice that bending rigidity and anticorrelation are both weaker in the passive polymer case with respect to the actively stretched one.

\subsection{Triple-helix torus knots $\mathcal{T}_3$}
\begin{figure}
\centering
\includegraphics[width=1.0\textwidth]{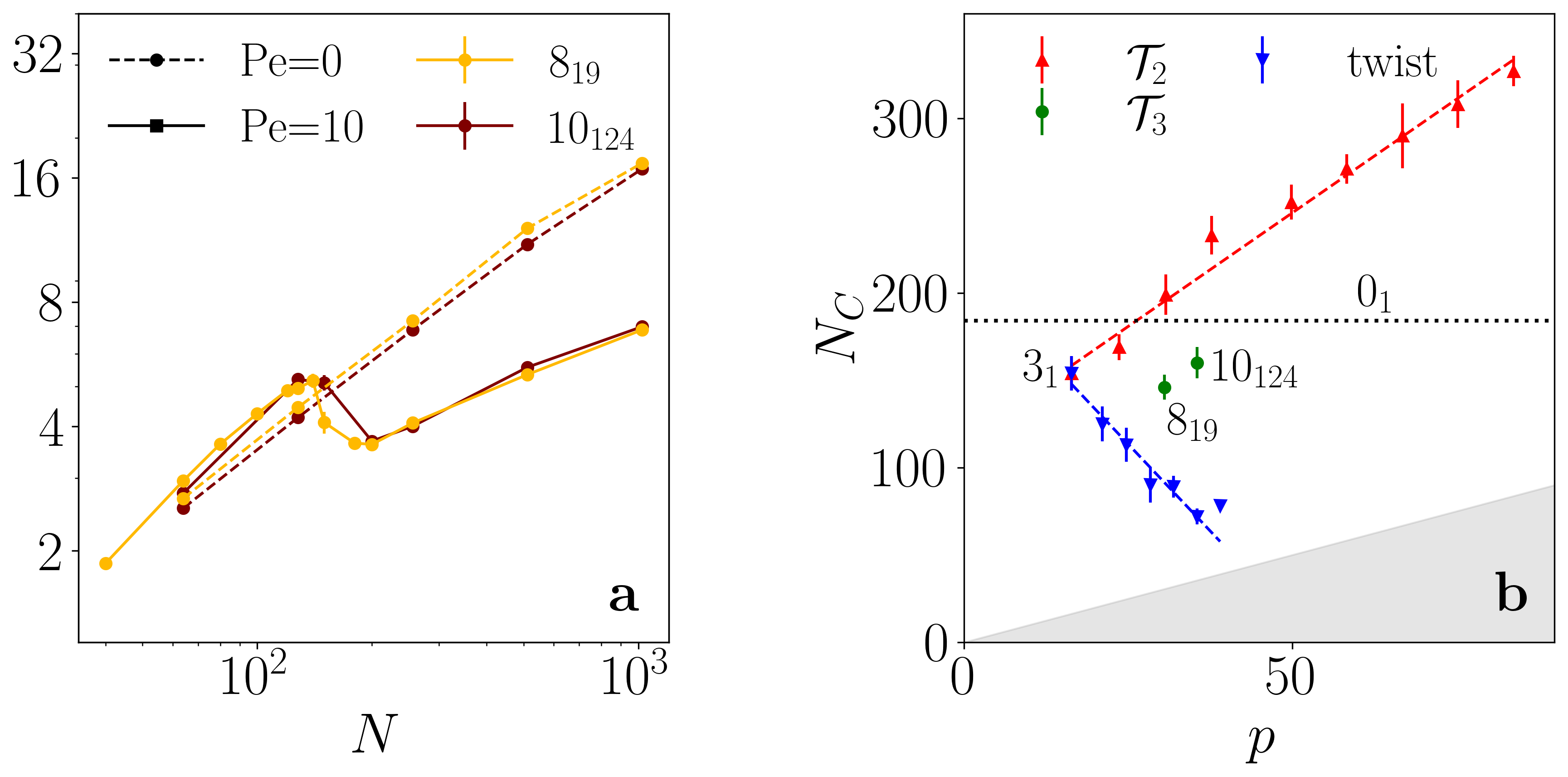}
\caption{(a) Gyration radius $R_g$ of passive (- -) and active(---) $\mathcal{T}_3$ knots, as a function of the polymer length $N$. (b) Collapse point $N_C$ of $\mathcal{T}_2$ (red upward triangles), $\mathcal{T}_3$ (green circles) and twist knots (blue downward triangles) as a function of the ideal length/diameter ratio $p$}
\label{fig:t3_collapse}
\end{figure}

\begin{figure}
\centering
\includegraphics[width=1.0\textwidth]{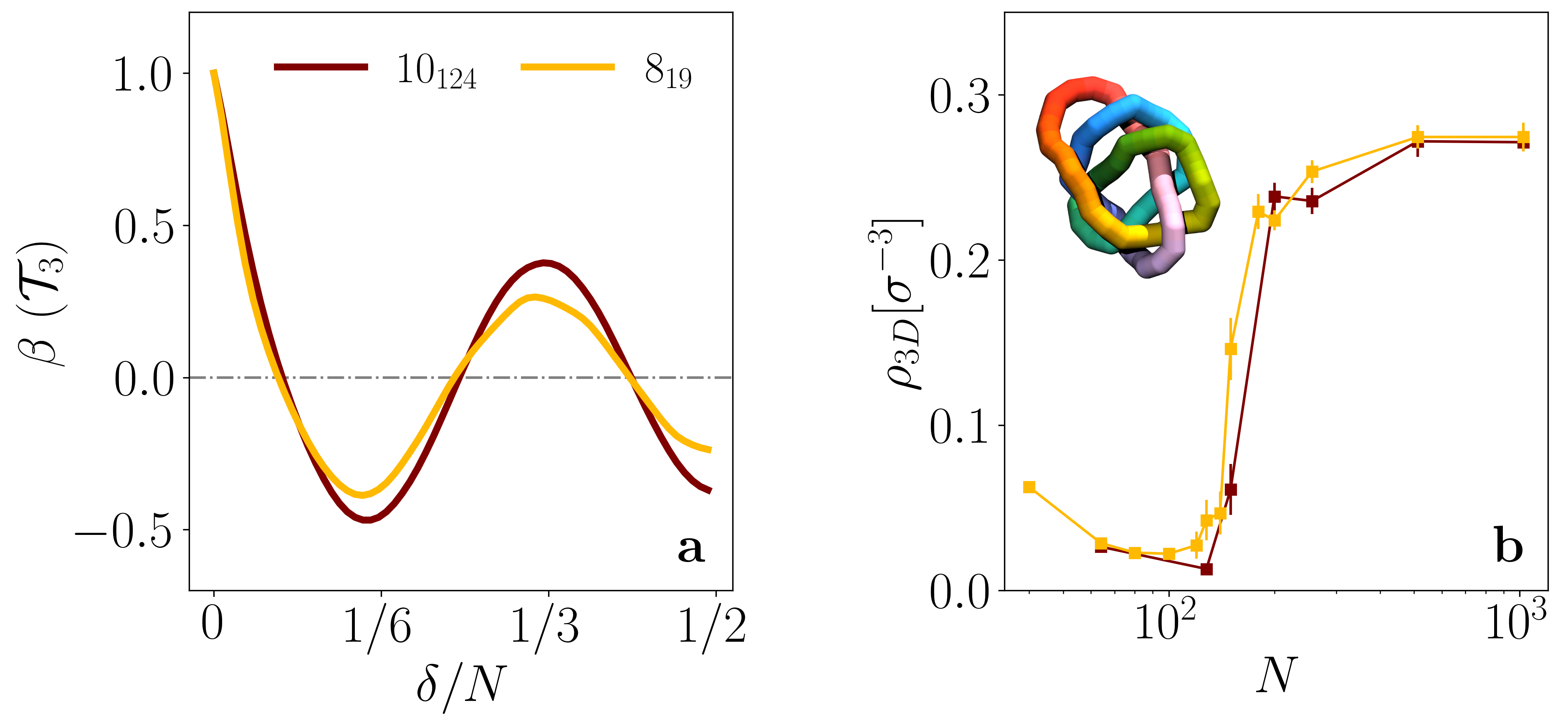}
\caption{(a) Bond correlation function $\beta(\delta)$ of active ring $\mathcal{T}_3$ polymers for $N=128$. (b) Number density of close bonds oppositely oriented $\rho_{\text{3D}}$ of active ring  $\mathcal{T}_3$ polymers as a function of the polymer length $N$. The inset is a simulation snapshot of an $8_{19}$ knot with $\text{Pe}=10$, $N=64$.}
\label{fig:t3_config}
\end{figure}

We studied the first two triple-helix torus knots $\mathcal{T}_3$ ($8_{19}$ and $10_{124}$), in order to assess whether they follow a behavior similar to that of $\mathcal{T}_2$ knots, and if so, to what extent. We see in Fig.~\ref{fig:t3_collapse} that both knots have a collapsing point close to that of the trefoil knot (Fig.~\ref{fig:t3_collapse}a,b), with a slight increase in the $N_C$ as $p$ increases, which is more compatible with the behavior of $\mathcal{T}_2$ knots than with that of twist knots. This similarity is reinforced by the very regular shape of their bond correlation function (Fig.~\ref{fig:t3_config}a), having for both knots a minimum at $N/6$, and by the rather small $\rho_{3D}$ in the stretched state (Fig.~\ref{fig:t3_config}b). As these preliminary results only consider the first two $\mathcal{T}_3$ knots though, no final conclusion should be made on their overall behavior.

\end{document}